\documentclass[aps,prd,showpacs,superscriptaddress,preprintnumbers,floatfix,
preprint]{revtex4-1}

\usepackage{amsfonts,amsmath,amssymb,amsthm}
\usepackage{bm}
\usepackage{braket}
\usepackage{cases}
\usepackage{color}
\usepackage{graphicx}
\usepackage{hyperref}
\usepackage{latexsym}
\usepackage{dcolumn,booktabs}

\newcommand{\mbp}{\mathbf{p}}

\newcommand{\rmGeV}{\mathrm{GeV}}
\newcommand{\rmMeV}{\mathrm{MeV}}

\begin{document}
\title{Masses of doubly heavy-quark baryons in an extended
  chromomagnetic model}

\author{Xin-Zhen Weng} 
\email{xzhweng@pku.edu.cn} 

\affiliation{School of Physics, Peking University, Beijing 100871,
  China}
\affiliation{State Key Laboratory of Nuclear Physics and Technology,
  Peking University, Beijing 100871, China}

\author{Xiao-Lin Chen} 
\email{chenxl@pku.edu.cn} 

\affiliation{School of Physics, Peking University, Beijing 100871,
  China}

\author{Wei-Zhen Deng} \email{dwz@pku.edu.cn} 

\affiliation{School of Physics, Peking University, Beijing 100871,
  China}
\affiliation{State Key Laboratory of Nuclear Physics and Technology, 
  Peking University, Beijing 100871, China}

\begin{abstract}

  We extend the chromomagnetic model by further considering the effect
  of color interaction.  The effective mass parameters between quark
  pairs ($m_{qq}$ or $m_{q\bar{q}}$) are introduced to account for both
  the effective quark masses and the color interaction between the two
  quarks. Using the experimental masses of hadrons, the quark pair
  parameters are determined between the light quark pairs and the
  light-heavy quark pairs.  Then the parameters of heavy quark pairs
  ($cc$, $cb$, $bb$) are estimated based on simple quark model
  assumption. We calculate all masses of doubly and triply
  heavy-quark baryons.  The newly discovered doubly charmed baryon
  $\Xi_{cc}$ fits into the model with an error of $12\,\rmMeV$.
\end{abstract}


\maketitle

\thispagestyle{empty}

\section{Introduction}\label{Sec:Introduction}

In 2002, the SELEX Collaboration~\cite{Mattson:2002vu} reported the
first observation of a doubly charmed baryon $\Xi_{cc}^{+}$ in the
decay mode $\Xi_{cc}^{+}\to\Lambda_c^{+}K^{-}\pi^{+}$.
Its mass was determined to be $3519\pm1\,\rmMeV$.
Further works identified its isospin partner $\Xi_{cc}^{++}(3460)$
\cite{Russ:2002bw} and an excited state
$\Xi_{cc}^{++}(3780)$~\cite{Moinester:2002uw}.
Later the SELEX Collaboration confirmed the $\Xi_{cc}^{+}$ state in
the $\Xi_{cc}^{+}\to{p}D^{+}K^{-}$ \cite{Ocherashvili:2004hi,
  Engelfried:2005kd} and
$\Xi_{cc}^{+}\to\Xi_{c}^{+}\pi^{+}\pi^{-}$~\cite{Engelfried:2007at}
decay modes.
However, none of these states were confirmed by other experimental
collaborations
\cite{Ratti:2003ez, Aubert:2006qw, Kato:2013ynr, Aaij:2013voa} 
in the subsequent searches.
Recently, the LHCb Collaboration~\cite{Aaij:2017ueg} reported the
observation of $\Xi_{cc}^{++}$ in the $\Lambda_c^+K^-\pi^+\pi^+$ decay
mode.
But its mass was determined to be $3621.40\pm0.72(\text{stat.})
\pm0.27(\text{syst.}) \pm0.14(\Lambda_c^+)\,\rmMeV$.

In contrast to the rarity of the experimental observation of the
doubly heavy baryons, there is a vast literature of theoretical
studies concerning the doubly and even triply heavy baryons with
different approaches, including
quark models \cite{DeRujula:1975qlm, Bjorken:1985ei, 
	SilvestreBrac:1996bg, Gershtein:2000nx, Kiselev:2002iy, 
	Ebert:2002ig, Karliner:2014gca, Migura:2006ep, Martynenko:2007je,
	Roberts:2007ni, Branz:2010pq,
	Valcarce:2008dr,
	Shah:2017jkr},
%
QCD sum rules~\cite{Zhang:2009re, Wang:2010hs, 
	Tang:2011fv, Aliev:2012iv, 
	Chen:2017sbg},
%
lattice QCD \cite{AliKhan:1999yb, Woloshyn:2000fe, Lewis:2001iz,
  Flynn:2003vz, Chiu:2005zc,
  Mehen:2006vv, Liu:2009jc, Meinel:2010pw,
  Alexandrou:2012xk, Briceno:2012wt, Durr:2012dw,
  Namekawa:2013vu, Padmanath:2015jea, Brown:2014ena, Vijande:2014uma,
  Bali:2015lka},
the bag model~\cite{Hasenfratz:1980ka}, 
heavy-quark effective theory~\cite{Korner:1994nh},
heavy-quark spin symmetry~\cite{Flynn:2011gf,
	Ma:2017nik},
effective field theory with potential nonrelativistic QCD~\cite{Brambilla:2005yk, LlanesEstrada:2011kc},
the Feynman-Hellmann theorem~\cite{Roncaglia:1995az}, 
variational method~\cite{Jia:2006gw}, 
the Skyrmion model~\cite{Rho:1990uy} 
and the Regge phenomenology
\cite{Burakovsky:1997vm, Guo:2008he}.

The quark model is one of the most used approaches to study the mass
spectra of hadrons \cite{Neeman:1961jhl, GellMann:1962xb,
  GellMann:1964nj, Zweig:1964jf,
  Eichten:1978tg, 
  Quigg:1977dd, 
  Martin:1980rm, 
  DeRujula:1975qlm, Isgur:1977ef, 
  Basdevant:1984rk, 
  Godfrey:1985xj, Capstick:1986bm, Godfrey:2004ya, Godfrey:2015dva,
  Godfrey:2016nwn}.
In the nonrelativistic limit, the QCD interaction can be reduced to
the potential interaction between quarks. Usually the potential
interaction in a quark model consists of the spin-independent color
interaction including the linear confinement and Coulomb-type terms,
plus higher order terms such as the spin-spin chromomagnetic
interaction, tensor interaction, and spin-orbit interactions.

When focusing on lowest $S$-wave states of hadrons, one may adopt the
chromomagnetic model~\cite{Sakharov:1966tua, Sakharov:1967,
  DeRujula:1975qlm, DeGrand:1975cf, Jaffe:1976ig, Jaffe:1976ih,
  SilvestreBrac:1992mv, Cui:2005az, 
  Buccella:2006fn, BorkaJovanovic:2010yc, Wu:2017weo}.
The chromomagnetic model assumes a mass formula by simply adding a term
of chromomagnetic hyperfine interaction to the effective quark masses.
This simplified model gives a good account of the hyperfine splittings
in hadron mass spectra and produces many useful Gell-Mann--Okubo (GMO)
mass relations.
From the point of view of the quark model, the effective quark masses also
include the chromoelectric effects from the color interaction.
However it is difficult to account for the two-body chromoelectric
effects in all relevant mesons and baryons by the effective quark
masses, which are one-body type. 
In Ref.~\cite{Hogaasen:2013nca},
H\o{}gaasen {\it et al.}  generalized the chromomagnetic model by
including a chromoelectric term
$H_{\text{CE}} = -\sum_{i,j} A_{ij}
\tilde{\bm\lambda}_i\cdot\tilde{\bm\lambda}_j$.
Similarly, Karliner {\it et al.} introduced the color-singlet binding
energies $B(c\bar{c})=-242.7\,\rmMeV$ and
$B(b\bar{b})=-532.2\,\rmMeV$~\cite{Karliner:2016zzc}.

In this paper, we use the extended chromomagnetic model with the
chromoelectric term to study the mass spectra of all the lowest
$S$-wave doubly and triply heavy-quark baryons systematically.
In Sec.~\ref{Sec:Model} we introduce the extended chromomagnetic
model and construct the model wave functions of mesons and baryons.
In Sec.~\ref{Sec:Parameter} we determine the model parameters.
The numerical results are presented and discussed in
Sec.~\ref{sec:mass-hh}.
We conclude in Sec.~\ref{Sec:Conclusion}.

\section{The Extended Chromomagnetic Model}\label{Sec:Model}

\subsection{The Hamiltonian}
\label{sec:hamiltonian}

In the quark model, the quark effective Hamiltonian reads
\cite{Godfrey:1985xj, Capstick:1986bm}
\begin{equation}
  H=H_0 + \sum_{i<j} V_{ij},
\end{equation}
where 
\begin{equation}
  H_0 = \sum_i \sqrt{\mbp_i^2+ m_i^2}
\end{equation}
is the relativistic mass term and $V_{ij}$ is the quark interaction
potential between $i$th and $j$th quarks.
In a nonrelativistic reduction,
\begin{equation}
  H_0 \to \sum_i \left( m_i + \frac{\mbp_i^2}{2m_i}\right),
\end{equation}
and
\begin{equation}
  V_{ij} \to V_{ij}^{\text{conf}} +  V_{ij}^{\text{hyp}} +  V_{ij}^{\text{so}},
\end{equation}
where
\begin{equation}\label{eqn:GI:conf}
  V_{ij}^{\text{conf}} = - \left[ \frac34 c + \frac34 br 
      - \frac{\alpha_s(r)}{r} \right] \bm{F}_i\cdot \bm{F}_j
\end{equation}
includes the color linear confinement and the Coulomb-type interaction,
$V_{ij}^{\text{hyp}}$ is the color hyperfine interaction, and
$V_{ij}^{\text{so}}$ is the spin-orbit interaction.
For the $S$-wave hadron, $V_{ij}^{\text{so}}$ has no contribution, 
and $V_{ij}^{\text{hyp}}$ can be simply replaced by 
the chromomagnetic interaction
\begin{equation}
  \label{cm-interaction}
  V_{ij}^{\text{cm}} = -\frac{8\pi}{3} \frac{\alpha_s(r)}{m_i m_j}
  \delta^3(\bm{r}) \bm{S}_i\cdot\bm{S}_j \bm{F}_i\cdot \bm{F}_j,
\end{equation}
where $\bm{S}_{i}$ and $\bm{F}_{i}$ are the $i$th quark's spin 
operator and color operator respectively,
\begin{equation}
  \bm{F}_i =
  \begin{cases}
    +\frac{\bm{\lambda}_i}2 & \text{for quarks}, \\
    -\frac{\bm{\lambda}_i^*}2 & \text{for antiquarks}.
  \end{cases}
\end{equation}

In the case of the lowest $S$-wave hadron, one may further simplify the
chromomagnetic interaction by ignoring its spatial dependency. Then
the chromomagnetic model Hamiltonian reads
\begin{equation}
H = \sum_{i} m_{i} + \sum_{i<j} V_{ij}^{\text{cm}} ,
\end{equation}
where the effective mass $m_i$ of $i$th constituent quark (or
antiquark) should include the constituent quark mass and the kinetic
energy and chromoelectric effects from $V_{ij}^{\text{conf}}$. The
chromomagnetic interaction reads
\begin{equation}
  V^{\text{cm}}_{ij} = - v_{ij} \bm{S}_{i}\cdot\bm{S}_{j} 
  \bm{F}_{i} \cdot \bm{F}_{j} \,.
\end{equation}
The coefficient $v_{ij}$ depends on the quark masses and the spatial
wave function of the hadron
\begin{equation}
  v_{ij}= \frac{8\pi}{3m_i m_j}
  \left\langle\alpha_s(r)\delta^3(\bm{r}) \right\rangle.
\end{equation}

However it is difficult to adsorb all the two-body chromoelectric
effects into the one-body effective quark masses if we want to study
all lowest $S$-wave mesons and baryons together
\cite{Hogaasen:2013nca, Karliner:2016zzc}. 
In Ref.~\cite{Hogaasen:2013nca}, H\o{}gaasen {\it et al.}  generalized
the chromomagnetic model by including a chromoelectric term
\begin{equation}
  H_{\text{CE}} = -\sum_{i,j} A_{ij} 
  \tilde{\bm\lambda}_i\cdot\tilde{\bm\lambda}_j,
\end{equation}
where $\tilde{\bm\lambda}_i=2\bm{F}_i$. We use this extended
chromomagnetic model to study all lowest $S$-wave mesons and baryons
systematically.

Since
\begin{align}\label{eqn:m+color=color}
&\sum_{i<j} \left( m_i + m_j \right) \bm{F}_{i} \cdot \bm{F}_{j} \notag \\
=& \frac12 \sum_{i,j} \left( m_i + m_j \right) \bm{F}_{i} \cdot \bm{F}_{j} 
- \sum_{i} m_i \bm{F}_i^2 \notag \\
=& \left(\sum_i m_i\bm{F}_i\right) \cdot \left(\sum_i \bm{F}_i\right) 
- \frac43 \sum_i m_i \,,
\end{align}
and the color operator $\sum_i\bm{F}_i$ nullifies any colorless
physical state,
we can introduce a new mass parameter of quark pair
\begin{equation}\label{eqn:para:color+m}
  m_{ij} = \left( m_i + m_j\right) + \frac{16}3 A_{ij} \,.
\end{equation}
Then the model Hamiltonian reads
\begin{equation}
  H_{\text{CM}} = -\frac34\sum_{i<j} m_{ij} V^{\text{C}}_{ij} 
  - \sum_{i<j} v_{ij} V^{\text{CM}}_{ij},
\end{equation}
where we have briefly introduced two operators to represent the color
and chromomagnetic (CM) interactions between quarks,
\begin{align}
  V^{\text{C}}_{ij} =& \bm{F}_{i} \cdot \bm{F}_{j} \,, \\
  V^{\text{CM}}_{ij} =& \bm{S}_{i}\cdot \bm{S}_{j} 
  \bm{F}_{i}^{a}\cdot\bm{F}_{j}^{a} \,.
\end{align}

For the mesons the Hamiltonian is simplified to
\begin{equation}
  H_{\text{CM}} = m_{q\bar{q}}
  - v_{q\bar{q}} V^{\text{CM}}_{q\bar{q}},
\end{equation}
and for the baryons
\begin{equation}
  H_{\text{CM}} = \frac12 \sum_{i<j} m_{ij}
  - \sum_{i<j} v_{ij} V^{\text{CM}}_{ij} .
\end{equation}
Since the quark model parameters of the baryon system usually are
different from that of the meson system, we assume that the pair
parameters $m_{qq}$ and $v_{qq}$ are different from their partners
$m_{q\bar{q}}$ and $v_{q\bar{q}}$ respectively. 
Their relations are studies in the next section, 
based on the numerical analysis and
the quark model consideration.

\subsection{Mesons}\label{sec::mesons}

A meson is a color-singlet hadron composed of a quark and an antiquark. 
Its total spin is either $1$ or $0$.
The corresponding spin wave functions are denoted by
\begin{align}
  \chi_{1m} &= \left\{ \ket{\uparrow\uparrow} \,, 
    \frac{1}{\sqrt{2}} \left( \ket{\uparrow\downarrow} 
      + \ket{\downarrow\uparrow} \right) \,, 
    \ket{\downarrow\downarrow} \right\} \,, \\
  \chi_{00} &= \frac{1}{\sqrt{2}} \left( \ket{\uparrow\downarrow}
    - \ket{\downarrow\uparrow} \right) \,,
\end{align}
where $m$ is the third component of the total spin.

The masses of the pseudoscalar and vector mesons are given by
\begin{align}
  M_{J=0} &= m_{q\bar{q}} - v_{q\bar{q}} \,, \\
  M_{J=1} &= m_{q\bar{q}} + \frac{1}{3} v_{q\bar{q}} \,.
\end{align}

\subsection{Baryons}
\label{sec::baryons}

Baryons are composed of three quarks. 
Since we only consider the lowest $S$-wave baryons
and the color wave function is antisymmetric,
we only have to construct the symmetric $\text{spin} \otimes
\text{flavor}$ wave functions.

The total spin of the baryon can be either $3/2$ or $1/2$. The spin
wave functions are classified according to the permutation symmetry,
\begin{align}
  \chi_{\frac32m}^{\text{S}} &= \left\{ \ket{\uparrow\uparrow\uparrow} \,, 
    \frac{1}{\sqrt{3}} \left( \ket{\uparrow\uparrow\downarrow}
      + \ket{\uparrow\downarrow\uparrow}
      + \ket{\downarrow\uparrow\uparrow} \right) \,, 
    \frac{1}{\sqrt{3}} \left( \ket{\uparrow\downarrow\downarrow}
      + \ket{\downarrow\uparrow\downarrow}
      + \ket{\downarrow\downarrow\uparrow} \right) \,, 
    \ket{\downarrow\downarrow\downarrow} \right\} \,, \\
  \chi_{\frac12m}^{\text{MS}} &= \left\{ - \frac{1}{\sqrt{6}} \left( 
      \ket{\uparrow\downarrow\uparrow}
      + \ket{\downarrow\uparrow\uparrow}
      - 2 \ket{\uparrow\uparrow\downarrow} \right) \,, 
    \frac{1}{\sqrt{6}} \left( \ket{\uparrow\downarrow\downarrow}
      + \ket{\downarrow\uparrow\downarrow}
      - 2 \ket{\downarrow\downarrow\uparrow} \right) \right\} \,, \\
  \chi_{\frac12m}^{\text{MA}} &= \left\{ \frac{1}{\sqrt{2}} \left( 
      \ket{\uparrow\downarrow\uparrow}
      - \ket{\downarrow\uparrow\uparrow} \right) \,, 
    \frac{1}{\sqrt{2}} \left( \ket{\uparrow\downarrow\downarrow}
      - \ket{\downarrow\uparrow\downarrow} \right) \right\} \,,
\end{align}
where the superscript MS (MA) suggests the symmetric (antisymmetric)
property of the wave functions under the exchange of the first two
quarks.

Next, we combine the flavor wave functions $\ket{q_1q_2q_3}$ with the
spin wave functions.  We get the following $\text{spin} \otimes
\text{flavor}$ base wave functions:
\begin{align}
  J=3/2: \quad & \phi_{\frac32m}^{\{q_1q_2q_3\}} = \ket{\{qqq\}} \otimes 
  \chi_{\frac32m}^{\text{S}} \,, \\
  J=1/2: \quad & \phi_{\frac12m}^{\{q_1q_2\}q_3} = \ket{\{q_1q_2\}q_3} \otimes 
  \chi_{\frac12m}^{\text{MS}} + \text{permutations}\,, \label{phi-1} \\
  & \phi_{\frac12m}^{[q_1q_2]q_3} = \ket{[q_1q_2]q_3} \otimes 
  \chi_{\frac12m}^{\text{MA}} + \text{permutations}\,, \label{phi-2}
\end{align}
where we use the brace $\{\cdots\}$ to symmetrize the quark flavors and
the bracket $\left[\cdots\right]$ to antisymmetrize the flavors.

The mass of the spin-$\frac32$ baryon is given by
\begin{equation}
  \label{mass-formula-32}
  M_{J=\frac32} = \frac12(m_{q_1q_2}+m_{q_1q_3} + m_{q_2q_3}) 
  + \frac16(v_{q_1q_2}+v_{q_1q_3} + v_{q_2q_3}).
\end{equation}
To obtain the masses of the spin-$\frac12$ baryons consisting of three
different quark flavors, we need to diagonalize the following
$2\times2$ matrix in the above basis [Eqs.~(\ref{phi-1}) and
(\ref{phi-2})],
\begin{equation}\label{eqn:hamitonian:baryon-1/2}
  H_{J=\frac12} = \frac12 (m_{q_1q_2}+m_{q_1q_3} + m_{q_2q_3}) +
  \begin{pmatrix}
    \frac{1}{6} \left( v_{q_1q_2} - 2 v_{q_1q_3} - 2v_{q_2q_3} \right)
    & - \frac{1}{2\sqrt{3}} 
    \left( v_{q_1q_3} - v_{q_2q_3} \right) \\
    - \frac{1}{2\sqrt{3}} \left( v_{q_1q_3} - v_{q_2q_3} \right) &
    -\frac{1}{2}v_{q_1q_2}
    \end{pmatrix} \,,
\end{equation}
which gives us two mixed states, which we denote by
$\phi_{\frac12m}^{q_1q_2q_3\pm}$, with masses
\begin{equation}
  M_{J=\frac12}^{\pm} = \frac12 (m_{q_1q_2}+m_{q_1q_3} + m_{q_2q_3}) 
  - \frac{1}{6} \left( v_{q_1q_2}+v_{q_1q_3} + v_{q_2q_3} \right)
  \pm \Delta_{J=\frac12},
\end{equation}
respectively,
where
\begin{equation}
  \Delta_{J=\frac12} = \frac13 \sqrt{v_{q_1q_2}^2+v_{q_1q_3}^2+v_{q_2q_3}^2 
    - v_{q_1q_2}v_{q_1q_3} - v_{q_1q_2}v_{q_2q_3} - v_{q_1q_3}v_{q_2q_3} }  \,.
\end{equation}

Note that if the flavors of any two quarks in the baryon are identical,
we can assign $q_1=q_2$ and the second combination [Eq.~(\ref{phi-2})]
does not exist. Then we get only one spin-$\frac12$ baryon state
$\phi^{\{q_1q_1\}q_3}_{\frac12 m}$ with mass
\begin{equation}
  \label{mass-formula-12}
  E_{J=\frac12}= \frac12 (m_{q_1q_1}+2m_{q_1q_3})+ \frac16(v_{q_1q_1}-4v_{q_1q_3}). 
\end{equation}

We collect the wave function assignments of all lowest $S$-wave
baryons in Table~\ref{table:hamiltonian-baryon}.
\begingroup
\squeezetable
\begin{table}
  \begin{ruledtabular}
    \caption{Baryon assignments.}
    \label{table:hamiltonian-baryon}
    \begin{tabular}{ccccc}
      Flavor & Spin-$\frac12$-baryon & Assignment & Spin-$\frac32$-baryon 
      & Assignment \\
      \hline
      $nnn$ & $N$ & $\phi_{\frac12m}^{\{nn\}n}$ & $\Delta$ & 
      $\phi_{\frac32m}^{\{nnn\}}$ \\
      $nns$ & $\Sigma$ & $\phi_{\frac12m}^{\{nn\}s}$ & $\Sigma^{*}$ & 
      $\phi_{\frac32m}^{\{nns\}}$ \\
      & $\Lambda$ & $\phi_{\frac12m}^{[nn]s}$ &  &  \\
      $nss$ & $\Xi$ & $\phi_{\frac12m}^{\{ss\}n}$ & $\Xi^*$ & 
      $\phi_{\frac12m}^{\{nss\}}$ \\
      $sss$ & & & $\Omega$ & $\phi_{\frac32m}^{\{sss\}}$ \\
      \hline
      $nnc$ & $\Sigma_c$ & $\phi_{\frac12m}^{\{nn\}c}$ & $\Sigma_c^*$ & 
      $\phi_{\frac32m}^{\{nnc\}}$ \\
      & $\Lambda_c$ & $\phi_{\frac12m}^{[nn]c}$ & &  \\
      $nsc$ & $\Xi_c$ & $\phi_{\frac12m}^{nsc-}$ & $\Xi_c^*$ & 
      $\phi_{\frac32m}^{\{nsc\}}$ \\
      & $\Xi_c'$ & $\phi_{\frac12m}^{nsc+}$ & & \\
      $ssc$ & $\Omega_c$ & $\phi_{\frac12m}^{\{ss\}c}$ & $\Omega_c^*$ & 
      $\phi_{\frac32m}^{\{ssc\}}$ \\
      \hline
      $nnb$ & $\Sigma_b$ & $\phi_{\frac12m}^{\{nn\}b}$ & $\Sigma_b^*$ & 
      $\phi_{\frac32m}^{\{nnb\}}$ \\
      & $\Lambda_b$ & $\phi_{\frac12m}^{[nn]b}$ & &  \\
      $nsb$ & $\Xi_b$ & $\phi_{\frac12m}^{nsb-}$ & $\Xi_b^*$ & 
      $\phi_{\frac32m}^{\{nsb\}}$ \\
      & $\Xi_b'$ & $\phi_{\frac12m}^{nsb+}$ & & \\
      $ssb$ & $\Omega_b$ & $\phi_{\frac12m}^{\{ss\}b}$ & $\Omega_b^*$ & 
      $\phi_{\frac32m}^{\{ssb\}}$ \\
      \hline\hline
      $ncc$ & $\Xi_{cc}$ & $\phi_{\frac12m}^{\{cc\}n}$ & $\Xi_{cc}^{*}$ & 
      $\phi_{\frac32m}^{\{ncc\}}$ \\
      $scc$ & $\Omega_{cc}$ & $\phi_{\frac12m}^{\{cc\}s}$ & $\Omega_{cc}^{*}$ & 
      $\phi_{\frac32m}^{\{scc\}}$ \\
      $ccc$ & & & $\Omega_{ccc}$ & $\phi_{\frac32m}^{\{ccc\}}$ \\
      \hline
      $nbb$ & $\Xi_{bb}$ & $\phi_{\frac12m}^{\{bb\}n}$ & $\Xi_{bb}^{*}$ & 
      $\phi_{\frac32m}^{\{nbb\}}$ \\
      $sbb$ & $\Omega_{bb}$ & $\phi_{\frac12m}^{\{bb\}s}$ & $\Omega_{bb}^{*}$ & 
      $\phi_{\frac32m}^{\{sbb\}}$ \\
      $bbb$ & & & $\Omega_{bbb}$ & $\phi_{\frac32m}^{\{bbb\}}$ \\
      \hline
      $ncb$ & $\Xi_{cb}$ & $\phi_{\frac12m}^{ncb-}$ & $\Xi_{cb}^*$ & 
      $\phi_{\frac32m}^{\{ncb\}}$ \\
      & $\Xi_{cb}'$ & $\phi_{\frac12m}^{ncb+}$ & & \\
      $scb$ & $\Omega_{cb}$ & $\phi_{\frac12m}^{scb-}$ & $\Omega_{cb}^*$ & 
      $\phi_{\frac32m}^{\{scb\}}$ \\
      & $\Omega_{cb}'$ & $\phi_{\frac12m}^{scb+}$ & & \\
      $ccb$ & $\Omega_{ccb}$ & $\phi_{\frac12m}^{\{cc\}b}$ & $\Omega_{ccb}^*$ &
      $\phi_{\frac32m}^{\{ccb\}}$ \\
      $cbb$ & $\Omega_{cbb}$ & $\phi_{\frac12m}^{\{bb\}c}$ & $\Omega_{cbb}^*$ & 
      $\phi_{\frac32m}^{\{cbb\}}$ \\
    \end{tabular}
  \end{ruledtabular}
\end{table}
\endgroup

\section{Numerical results}
\label{Sec:Result}

\subsection{Parameters}
\label{Sec:Parameter}

First we consider the mesons.
We can extract the two parameters $m_{q_1\bar{q}_2}$ and
$v_{q_1\bar{q}_2}$ from the experimental masses of corresponding
$q_1\bar{q}_2$ pseudoscalar and vector mesons.
For the $n\bar{n}$ mesons consisting of $u,d$ flavors, we only use the
isovector $\pi$ and $\rho$ mesons to extract $m_{n\bar{n}}$ and
$v_{n\bar{n}}$.

We do not consider $\eta$ and $\eta'$ mesons to avoid the complexity
of flavor octet-singlet mixing and the chiral anomaly.
Instead, we use the following PCAC (partially conserved axial current) result
\cite{Scadron:1982eg, Feldmann:1998su, Kroll:2004rs},
\begin{equation}\label{eqn:ss}
M_{s\bar{s}(^1S_0)} = \sqrt{2M_K^2-M_{\pi}^2} = 687.220\,\rmMeV \,,
\end{equation}
and the experimental mass of the $\phi$ meson to extract the parameters
$m_{s\bar{s}}$ and $v_{s\bar{s}}$.
The equation can also be derived in the chiral perturbation
theory~\cite{Scherer:2012xha}.

Another difficulty is that only one of the two $c\bar{b}$ states, that
is, the $B_c$ meson, was observed in experiment.
This state was first reported by CDF and OPAL collaborations in
1998~\cite{Abe:1998wi,Ackerstaff:1998zf}, whose current mass in PDG is
$6275.1\,\rmMeV$~\cite{Patrignani:2016xqp}.
Godfrey {\it et al.} had predicted its mass to be $6.27\,\rmGeV$ using
the quark model in 1985~\cite{Godfrey:1985xj}; a more detailed study in
2004 gives $M_{B_c}=6271\,\rmMeV$~\cite{Godfrey:2004ya}, which is very
closed to the experimental value.
They also predicted $M_{B_c^*}=6338\,\rmMeV$.
Other quark model calculation coincides with their result.
For instance, Ikhdair {\it et al.}~\cite{Ikhdair:2003ry} predict
$M_{B_c^*}=6340\,\rmMeV$ and Ebert {\it et al.}~\cite{Ebert:2002pp}
predict $M_{B_c^*}=6332\,\rmMeV$.
In our work, we use the prediction $M_{B_c^*}=6338\,\rmMeV$ of
Godfrey {\it et al.} to determine the parameters of the $c\bar{b}$ pair.
All the $q\bar{q}$ pair parameters are presented in
Table~\ref{table:parameter-meson}.
%
\begin{table}
  \begin{ruledtabular}
    \caption{Parameters of $q\bar{q}$ pairs 
      (in units of $\rmMeV$).}
    \label{table:parameter-meson}
    \begin{tabular}{cccccccccccc}
      $m_{n\bar{n}}$  & $m_{n\bar{s}}$ & $m_{s\bar{s}}$ 
      & $m_{n\bar{c}}$ & $m_{s\bar{c}}$ & $m_{c\bar{c}}$ 
      & $m_{n\bar{b}}$ & $m_{s\bar{b}}$ & $m_{c\bar{b}}$ 
      & $m_{b\bar{b}}$ \\
      $615.95$ & $794.22$ & $936.40$ & $1973.22$ 
      & $2076.14$ & $3068.53$ & $5313.35$ & $5403.25$ & $6322.27$ 
      & $9444.97$ \\
      \hline
      $v_{n\bar{n}}$  & $v_{n\bar{s}}$ & $v_{s\bar{s}}$ 
      & $v_{n\bar{c}}$ & $v_{s\bar{c}}$ & $v_{c\bar{c}}$ 
      & $v_{n\bar{b}}$ & $v_{s\bar{b}}$ & $v_{c\bar{b}}$ 
      & $v_{b\bar{b}}$ \\
      $477.92$ & $298.57$ & $249.18$ & $106.01$ & $107.87$ 
      & $85.12$ & $33.89$ & $36.43$ & $47.18$ & $45.98$ \\
    \end{tabular}
  \end{ruledtabular}
\end{table}
%


Now we turn to the baryon sector.
We can only use the experimental masses of light-quark baryons and
singly heavy-quark baryons to extract the model parameters.
Besides, the $\Omega_b^*$ has not yet been observed in experiment.
We perform an unweighted nonlinear least-squares fit of $23$ known
baryon masses to extract 13 model parameters, using the \texttt{GSL}
library~\cite{Galassi-2017-GNUScientificLibrary}. Note that, with two
identical quarks, the pair parameters $m_{qq}$ and $v_{qq}$ only
appear in the combination $m_{qq}+v_{qq}/3$ in the mass formulas
(\ref{mass-formula-12}) and (\ref{mass-formula-32}).  So we can only
determine the value $m_{ss}+v_{ss}/3$ from the experimental data.  

The baryon parameters obtained are presented in
Table~\ref{table:parameter-baryon}.
%
\begin{table}
  \begin{ruledtabular}
    \caption{Parameters of light-light and light-heavy quark pairs with
      statistical errors
      (in units of $\rmMeV$).}
    \label{table:parameter-baryon}
    \begin{tabular}{cccccccccccc}
      $m_{nn}$ & $m_{ns}$ & $m_{nc}$ & $m_{sc}$ & $m_{nb}$ 
      & $m_{sb}$ \\
      $724.85\pm3.37$ & $906.65\pm3.43$ & $2079.96\pm4.47$ 
      & $2183.68\pm5.33$ & $5412.25\pm4.81$ & $5494.80\pm10.05$ \\
      \hline
      $v_{n{n}}$  & $v_{n{s}}$ & $v_{n{c}}$ & $v_{s{c}}$ 
      & $v_{n{b}}$ & $v_{s{b}}$ \\
      $305.34\pm6.54$ & $212.75\pm6.06$ & $62.81\pm9.68$ 
      & $70.63\pm9.92$ & $19.92\pm10.19$ & $8.47\pm16.66$ \\
      \hline
      $m_{ss}+v_{ss}/3$ \\
      $1114.45\pm4.55$ \\
    \end{tabular}
  \end{ruledtabular}
\end{table}
The fitting standard deviation is $7.66\,\rmMeV$. Because the
$\Omega_b^*$ has not yet been observed in experiment, the parameters
$m_{sb}$ and $v_{sb}$ have large statistical errors.
The comparison of the fitted mass values with experimental data is
listed in Table~\ref{table:fitting-baryon}.
Most fitting deviations of the baryon masses are within $10\,\rmMeV$. 
The only exception is the $\Sigma$, whose deviation is $15.0\,\rmMeV$.
%
\begin{table}
  \begin{ruledtabular}
    \caption{Comparison for light and singly heavy-quark baryon masses
      (with statistical errors) with experimental
      data~\cite{Patrignani:2016xqp}  (in units of $\rmMeV$).}
    \label{table:fitting-baryon}
    \begin{tabular}{c|cccccccccc|ccccc|c}
      & $nnn$ & $nns$ & $nns$ & $nss$ & $sss$ \\
      \hline
      $J^P=1/2^+$ & $N(938.9)$ & $\Sigma(1193.2)$ & $\Lambda(1115.7)$ 
      & $\Xi(1318.3)$ & $-$ \\
      Theo. & $934.6\pm6.0$ & $1178.1\pm5.7$ & $1116.4\pm5.0$ 
      & $1322.1\pm5.8$ \\
      \hline
      $J^P=3/2^+$ & $\Delta(1232)$ & $\Sigma^*(1384.6)$ & $-$ & $\Xi^*(1533.4)$
      & $\Omega(1672.5)$ \\
      Theo. & $1239.9\pm6.0$ & $1390.9\pm4.5$ && $1534.8\pm4.6$ 
      & $1671.7\pm6.8$ \\
      \hline\hline
      & $nnc$ & $nnc$ & $nsc$ & $nsc$ & $ssc$ \\
      \hline
      $J^P=1/2^+$ & $\Sigma_c(2453.6)$ & $\Lambda_c(2286.5)$ 
      & $\Xi_c'(2576.8)$ & $\Xi_c(2469.4)$ & $\Omega_c(2695.2)$ \\
      Theo. & $2451.4\pm8.1$ & $2289.7\pm5.8$ & $2576.2\pm5.6$ 
      & $2478.7\pm5.6$ & $2693.8\pm8.8$ \\
      \hline
      $J^P=3/2^+$ & $\Sigma_c^{*}(2518.1)$ & $-$ & $\Xi_c^*(2645.9)$ 
      & $-$ & $\Omega_c^*(2765.9)$ \\
      Theo. & $2514.2\pm5.9$ && $2642.8\pm4.6$ && $2764.5\pm6.7$ \\
      \hline\hline
      & $nnb$ & $nnb$ & $nsb$ & $nsb$ & $ssb$ \\
      \hline
      $J^P=1/2^+$ & $\Sigma_b(5813.4)$ & $\Lambda_b(5619.5)$ & 
      $\Xi_b'(5935.0)$ & $\Xi_b(5793.2)$ & $\Omega_b(6046.4)$ \\
      Theo. & $5812.3\pm8.6$ & $5622.0\pm6.1$ & $5932.9\pm7.8$ 
      & $5800.4\pm7.8$ & $6046.4\pm15.1$ \\
      \hline
      $J^P=3/2^+$ & $\Sigma_b^{*}(5833.6)$ & $-$ & $\Xi_b^*(5952.1)$ 
      & $-$ & $\Omega_b^{*}$ \\
      Theo. & $5832.2\pm6.2$ && $5947.0\pm6.8$ && $6054.8\pm11.7$ \\
    \end{tabular}
  \end{ruledtabular}
\end{table}

In our model, all chromoelectric effects of color interaction are
included in the pair mass parameter $m_{qq}$ (or $m_{q\bar{q}}$).  If
the chromoelectric effects can be absorbed into the quark mass $m_q$
like in the original chromomagnetic model, we have the relation
\begin{equation*}
  m_{q_1\bar{q}_1}+m_{q_2\bar{q}_2} - 2m_{q_1\bar{q}_2}\approx 0.
\end{equation*}
This is not true from our fitting. Typically
\begin{equation*}
  m_{n\bar{n}}+m_{b\bar{b}} - 2m_{n\bar{b}} \approx -600 \,\rmMeV.
\end{equation*}

%
\begin{table}
  \begin{ruledtabular}
    \caption{Difference of pair mass parameters extracted from baryons
      and mesons (in units of $\rmMeV$).}
    \label{table:para:color:mass-diff:sum}
    \begin{tabular}{cccccc}
      $\delta m_{nn}$ & $\delta m_{ns}$ & $\delta m_{nc}$ & $\delta m_{sc}$ 
      & $\delta m_{nb}$ & $\delta m_{sb}$ \\
      \hline
      $108.89\pm3.37$ & $112.44\pm3.43$ & $106.74\pm4.47$ 
      & $107.54\pm5.33$ & $98.90\pm4.81$ & $91.54\pm10.05$ \\
    \end{tabular}
  \end{ruledtabular}
\end{table}
We also note that the quark pair mass $m_{qq}$ is quite different from
$m_{q\bar{q}}$ of its quark antiquark partner.
We list the difference 
$\delta{m_{q_1q_2}}\equiv{m_{q_1q_2}}-{m_{q_1\bar{q}_2}}$ 
in Table~\ref{table:para:color:mass-diff:sum}.
Indeed, many authors found that the effective quark masses extracted from
baryons were larger than that from mesons \cite{Karliner:2014gca,
  Karliner:2016zzc, Gasiorowicz:1981jz, Buccella:2006fn}.
This mass difference can be also accounted by adjusting the constant
$c$ in the quark interaction [Eq.~(\ref{eqn:GI:conf})], if it can be
treated as a constant~\cite{Godfrey:1985xj,Capstick:1986bm}.
Here we assume that
\begin{equation}
  A_{q\bar{q}} \approx A_{qq},
\end{equation}
in Eq.~(\ref{eqn:para:color+m})
and the difference of the pair mass parameter becomes
\begin{equation}\label{eqn:para:mass-diff}
  \delta m_{q_1q_2} \equiv m_{q_1q_2} - m_{q_1\bar{q}_2} 
  \approx \delta m_{q_1} + \delta m_{q_2} \,,
\end{equation}
where $\delta m_{q} = m_q^b - m_q^m$ is the difference of the effective
quark mass extracted from the baryon and meson.
Then we perform a least-squares fitting to obtain the mass difference
$\delta{m_q}$,
which is listed in Table~\ref{table:parameter-delta_m}.
%
\begin{table}
  \begin{ruledtabular}
    \caption{Quark mass difference $\delta m_q$ (in units of
      $\rmMeV$).}
    \label{table:parameter-delta_m}
    \begin{tabular}{cccccccccccc}
      $\delta{m_n}$  & $\delta{m_s}$ & $\delta{m_c}$ 
      & $\delta{m_b}$ \\
      \hline
      $54.94\pm1.51$ & $56.48\pm3.06$ & $51.49\pm3.68$ 
      & $42.30\pm4.51$ \\
    \end{tabular}
  \end{ruledtabular}
\end{table}
The reduced chi-squared statistic is $\chi_\nu^2=0.41$.

In Table~\ref{table:ratio-cm}, we compare the chromomagnetic
interaction strengths in baryons and mesons using their ratio
$R_{q_1q_2}\equiv v_{q_1q_2}/v_{q_1\bar{q}_2}$.
%
\begin{table}
  \begin{ruledtabular}
    \caption{Ratio of CM interaction strength
      $R_{q_1q_2}=v_{q_1q_2}/v_{q_1\bar{q}_2}$.}
    \label{table:ratio-cm}
    \begin{tabular}{c|cccccccccc}
      $q_1q_2$ & $nn$ & $ns$ & $nc$ & $sc$ & $nb$ & $sb$ \\
      \hline
      Ratio & $0.64\pm0.01$ & $0.71\pm0.02$ & $0.59\pm0.09$ 
      & $0.65\pm0.09$ & $0.59\pm0.30$ & $0.23\pm0.46$ \\
    \end{tabular}
  \end{ruledtabular}
\end{table}
We find that $R_{nn}$, $R_{ns}$, $R_{nc}$, $R_{sc}$ are very close to
each others. $R_{sb}$ is relatively small but with large statistical
error due to the lack of experimental data of $B_c^*$.
This phenomenon was first observed by
Keren-Zur~\cite{KerenZur:2007vp}. The ratio was interpreted in the quark
model, using the Cornell potential or the Logarithmic potential. The
author also gave a simple interpretation by assuming that the contact
probability in the chromomagnetic interaction
[Eq.~(\ref{cm-interaction})] is inversely proportional to the number
of quarks in the hadron.  Since the quark number is $3$ in a baryon
and $2$ in a meson, this gives a rough estimate of
$R_{q_1q_2} \approx 2/3$.
To estimate the heavy quark pair parameters $\{v_{cc} \,, v_{cb} \,,
v_{bb}\}$, we assume that
\begin{equation}\label{eqn:para:cm-ration}
R_{Q_1Q_2}=2/3\pm0.30 \,,
\end{equation}
where we use the largest statistical error in
Table~\ref{table:ratio-cm} (except $R_{sb}$ whose statistical error is
mainly due to the lack of experimental data) to set the parameter
range.
We should point out that even the estimate causes large standard errors
in $\{v_{cc},v_{cb},v_{bb}\}$; it does not have so many significant
effects on the mass of doubly and triply heavy-quark baryons as the
absolute values $v_{Q_1Q_2}$ are much smaller than $v_{qQ}$ between
light and heavy quarks.

Using the mass difference Eq.~(\ref{eqn:para:mass-diff}) and ratio
relation Eq.~(\ref{eqn:para:cm-ration}), we can determine the
parameters between two heavy quarks, as well as $m_{ss}$ and $v_{ss}$.
All the baryon parameters are collected in 
Table~\ref{table:parameter-baryon-all}.
%
\begin{table}
  \begin{ruledtabular}
    \caption{Parameters of $qq$ pairs
      (in units of $\rmMeV$).}
    \label{table:parameter-baryon-all}
    \begin{tabular}{cccccccccccc}
      $m_{nn}$ & $m_{ns}$ &  $m_{ss}$ & $m_{nc}$ 
      & $m_{sc}$ \\
      $724.85\pm3.37$ & $906.65\pm3.43$ & $1049.36\pm4.32$ 
      & $2079.96\pm4.47$ & $2183.68\pm5.33$ \\
      \hline
      $m_{cc}$ & $m_{nb}$ & $m_{sb}$ & $m_{cb}$ 
      & $m_{b{b}}$ \\
      $3171.51\pm5.21$ & $5412.25\pm4.81$ & $5494.80\pm10.05$ 
      & $6416.07\pm5.82$ & $9529.57\pm6.37$ \\
      \hline
      $v_{n{n}}$  & $v_{n{s}}$ & $v_{ss}$ & $v_{n{c}}$ 
      & $v_{s{c}}$ \\
      $305.34\pm6.54$ & $212.75\pm6.06$ & $195.30\pm18.84$ 
      & $62.81\pm9.68$ & $70.63\pm9.92$ \\
      \hline
      $v_{c{c}}$ & $v_{n{b}}$ & $v_{s{b}}$ & $v_{c{b}}$ 
      & $v_{b{b}}$  \\
      $56.75\pm25.54$ & $19.92\pm10.19$ & $8.47\pm16.66$ 
      & $31.45\pm14.15$ & $30.65\pm13.79$ \\
    \end{tabular}
  \end{ruledtabular}
\end{table}
%


\subsection{Mass spectra of doubly and triply heavy baryons}
\label{sec:mass-hh}

Substituting the parameters obtained in Sec.~\ref{Sec:Parameter} into
the Hamiltonians, we can obtain the masses of doubly and triply
heavy-quark baryons.
They are summarized in Table~\ref{table:mass-baryon-cc-bb}.
\begin{table}
  \begin{ruledtabular}
    \caption{Mass of the doubly and triply heavy baryons (in units of
      $\rmMeV$).}
    \label{table:mass-baryon-cc-bb}
    \begin{tabular}{c|cccccc}
      $$ & $ncc$ & $scc$ & $ccc$ & $nbb$ & $sbb$ & $bbb$ \\
      \hline
      $J^P=1/2^+$ & $\Xi_{cc}$ & $\Omega_{cc}^{+}$ & $-$ & $\Xi_{bb}$ 
      & $\Omega_{bb}^{-}$ & $-$ \\
      Exp. & $3518.7\pm1.7${\footnote{SELEX~\cite{Mattson:2002vu}.}} \\
      Exp. & $3621.40\pm0.72${\footnote{LHCb~\cite{Aaij:2017ueg}.}} \\
      Theo. & $3633.3\pm9.3$ & $3731.8\pm9.8$ && $10168.9\pm9.2$ 
      & $10259.0\pm15.5$ \\
      \hline
      $J^P=3/2^+$ & $\Xi_{cc}^{*}$ & $\Omega_{cc}^{*+}$ 
      & $\Omega_{ccc}^{*++}$ & $\Xi_{bb}^{*}$ & $\Omega_{bb}^{*-}$ 
      & $\Omega_{bbb}^{*-}$ \\
      Theo.  & $3696.1\pm7.4$ & $3802.4\pm8.0$ & $4785.6\pm15.0$ 
      & $10188.8\pm7.1$ & $10267.5\pm12.1$ & $14309.7\pm11.8$ \\
      \hline\hline
      $$ & \multicolumn{2}{c}{$ncb$} & \multicolumn{2}{c}{$scb$} & $ccb$ 
      & $cbb$ \\
      \hline
      $J^P=1/2^+$ & $\Xi_{cb}'$ & $\Xi_{cb}$ & $\Omega_{cb}'^{0}$ 
      & $\Omega_{cb}^{0}$ & $\Omega_{ccb}^{+}$ & $\Omega_{cbb}^{0}$ \\
      Theo.  & $6947.9\pm6.9$ & $6922.3\pm6.9$ & $7047.0\pm9.3$ 
      & $7010.7\pm9.3$ & $7990.3\pm12.2$ & $11165.0\pm11.8$ \\
      \hline
      $J^P=3/2^+$ & $\Xi_{cb}^{*}$ &  & $\Omega_{cb}'^{*0}$ & 
      & $\Omega_{ccb}^{*+}$ & $\Omega_{cbb}^{*0}$ \\
      Theo.  & $6973.2\pm5.5$ && $7065.7\pm7.5$ && $8021.8\pm9.0$ 
      & $11196.4\pm8.5$ \\
    \end{tabular}
  \end{ruledtabular}
\end{table}
In our calculation $M_{\Xi_{cc}}=3633.3\pm9.3\,\rmMeV$.
It is much heavier than the SELEX's value by approximately 
$100\,\text{MeV}$~\cite{Mattson:2002vu}, and very closed to the report of 
LHCb~\cite{Aaij:2017ueg}.
The $\Xi_{cc}^*$ state lies $62.8\,\rmMeV$ above $\Xi_{cc}$.
This splitting is very closed to the one between ${\Sigma_c}$ and 
${\Sigma_c^*}$ ($64.5\,\rmMeV$), which is consistent with the GMO mass 
relation~\cite{Johnson:1976is}
\begin{equation}
M_{\Xi_{cc}^*}-M_{\Xi_{cc}} = M_{\Sigma_c^*}-M_{\Sigma_c} \,.
\end{equation}
A similar relation holds if we replace the $u$, $d$ quarks by the $s$ quark
\begin{equation}
M_{\Omega_{cc}^*}-M_{\Omega_{cc}} = M_{\Omega_c^*}-M_{\Omega_c} \,,
\end{equation}
where both sides are approximately $71\,\rmMeV$.
Similar to the $\Sigma_c^{(*)}$ (or $\Omega_{c}^{(*)}$) case, the splitting 
between $\Xi_{cc}^*$ and $\Xi_{cc}$ (or between $\Omega_{cc}^*$ and 
$\Omega_{cc}$) is too small to induce a transition through the emission 
of the $\pi$ meson; however, the transition is still possible through $\gamma$ 
emission.

The situation for bottomed baryons is similar;
\begin{equation}
M_{\Xi_{bb}^*}-M_{\Xi_{cc}} \approx M_{\Sigma_b^*}-M_{\Sigma_b} \,,
\end{equation}
where the left-hand side is $19.9\,\rmMeV$ and the right-hand side 
is $20.2\,\rmMeV$.
This splitting is significantly smaller than that of charmed baryons.
The reason is that the hyperfine splitting is reciprocal to the masses 
of quarks, and of course the $b$ quark is much heavier than the $c$ quark.

There is also one GMO mass relation about the triply heavy-quark baryons,
that is
\begin{equation}
M_{\Omega_{cbb}^{*}}-M_{\Omega_{cbb}} = 
M_{\Omega_{ccb}^{*}}-M_{\Omega_{ccb}}
\end{equation}
where both sides are approximately $31\,\rmMeV$.

For spin-$1/2$ doubly heavy-quark baryons composed of three different
quarks, namely the $qcb$ baryon states ($q=u,d,s$), one should consider
the mixture between two basis states (\ref{phi-1}) and (\ref{phi-2}). 
Numerically, the mixing matrix in Eq.~(\ref{eqn:hamitonian:baryon-1/2})
is given by (in $\rmMeV$)
\begin{equation*}
\begin{pmatrix}
-6.7 & 3.3 \\
3.3 & -31.4
\end{pmatrix}
\quad \text{and} \quad
\begin{pmatrix}
-1.5 & 6.6 \\
6.6 & -35.3
\end{pmatrix} \,,
\end{equation*}
for $ncb$ and $scb$ flavor configurations respectively.
The eigenvalues of $ncb$ states are $\{-31.8,-6.2\}$, with 
eigenvectors $\{-0.13,0.99\}$ and $\{0.99,0.13\}$, 
and the eigenvalues of $scb$ states are $\{-36.6,-0.3\}$ with
eigenvectors $\{-0.19,0.98\}$ and $\{0.98,0.19\}$.
In both cases, the mixing is very small and the mixing mass effects
are within $2\,\rmMeV$.

If one ignores the mixing, then the $\Xi_{cb}$ and $\Xi_{cb}'$ can be
treated as states in the flavor $SU(2)_{nc}$ singlet and triplet
representations and the $\Omega_{cb}$ and $\Omega_{cb}'$ as states in
the $SU(2)_{sc}$ singlet and triplet representations, respectively
\cite{Roberts:2007ni}. The following GMO mass relations hold
approximately:
\begin{align}
& 2M_{\Xi_{cb}^*}+M_{\Xi_{cb}'}-3M_{\Xi_{cb}} \approx 
2\left(M_{\Sigma_c^*}-M_{\Sigma_c}\right) \,, \\
& 2M_{\Omega_{cb}^*}+M_{\Omega_{cb}'}-3M_{\Omega_{cb}} \approx 
2\left(M_{\Omega_c^*}-M_{\Omega_c}\right) \,, \\
& 2\left(M_{\Xi_{cb}^{*}}-M_{\Xi_{cb}'}\right) - 
\left(M_{\Omega_{cbb}^{*}}-M_{\Omega_{cbb}}\right) \approx 
M_{\Sigma_{b}^{*}}-M_{\Sigma_{b}} \,, \\
& 2\left(M_{\Omega_{cb}^{*}}-M_{\Omega_{cb}'}\right) - 
\left(M_{\Omega_{cbb}^{*}}-M_{\Omega_{cbb}}\right) \approx 
M_{\Omega_{b}^{*}}-M_{\Omega_{b}} \label{eqn:GMO:mixing:4} \,.
\end{align}
We find that the errors of all those relations are within $5\,\rmMeV$.

\section{Conclusions}
\label{Sec:Conclusion}

In this work, we generalized the chromomagnetic model by considering
the effect of color interaction.
According to color algebra, the quark effective mass and the color
interaction between quarks are combined into a new quark pair mass
parameter.
The quark pair parameters between two light quarks and that between
light-heavy quarks are determined using the experimental masses
of lowest $S$-wave hadrons.
The pair parameters between two heavy quarks are estimated from the
corresponding pair parameters between the quark and antiquark in mesons,
using the mass difference and a ratio relation about the
chromomagnetic interaction.
We have calculated the mass spectra of the lowest $S$-wave doubly and
triply heavy baryons.
We obtained $M_{\Xi_{cc}}=3633.3\pm9.3\,\rmMeV$, which is close
to the report of LHCb.
We hope that future experiments in LHCb, BES-III {\it et al.}
confirm the existence of these states.

\section*{ACKNOWLEDGMENTS}

The authors thank S.~L.~Zhu for helpful comments and discussions.
X.~Z.~W. is grateful to J.~F.~Jiang, H.~S.~Li, L.~Meng, G.~J.~Wang,
S.~J.~Wu, and B.~Zhou for helpful discussions.
This project is supported by the National Natural Science Foundation
of China under Grants No. 11621131001 and 11575008.

\bibliography{myreference}
\end{document}